\documentclass[11pt]{article}
\normalsize
\usepackage{amsfonts}
\usepackage{amsmath}
\usepackage{amssymb}
\usepackage{latexsym}
\usepackage{enumerate}

\newcommand{\F}{\mathbb{F}}
\newcommand{\N}{\mathbb{N}}

\newcommand{\ho}{\mbox{\rm Hom}}

\newcommand{\wt}{\mbox{\rm wt}}

\newcommand{\dist}{\mbox{\rm d}}

\title{\bf A Note on Linear Complementary Pairs of Group Codes }
\author{}

\author{Martino Borello, \\ LAGA, UMR 7539, CNRS, Universit\'e Paris 13 - Sorbonne Paris Cit\'e,\\ Universit\'e Paris 8, F-93526,
Saint-Denis, France.
 \\ and \\
Javier de la Cruz \\Universidad del Norte, Barranquilla, Colombia\\
and \\ Wolfgang Willems \\ Otto-von-Guericke Universit\"at, Magdeburg, Germany  \\
 and Universidad del Norte, Barranquilla, Colombia
}

\date{}

\begin{document}

\maketitle


\begin{abstract} We give a short and elementary proof of the fact that
for a linear complementary pair $(C,D)$,
where $C$ and $D$ are $2$-sided ideals in a group algebra, $D$ is
uniquely determined by $C$ and the dual code $D^\perp$ is
permutation equivalent to $C$. This includes earlier results of
\cite{C18} and \cite{G18} on nD cyclic codes which have been proved
by subtle and lengthy calculations in the space of polynomials.
\end{abstract}

\mbox{}\\

Throughout this note let $K$ be a finite field. A pair $(C,D)$ of linear codes over $K$ of length $n$
is called a {\it \underline{l}inear \underline{c}omplementary \underline{p}air} (LCP) if $C \cap D = \{0\}$ and $C +D = K^n$,
or equivalently if $C \oplus D = K^n$. In the special case that $D=C^\perp$  where the dual is taken
with respect to  the Euclidean inner product the code $C$ is referred to a
{\it \underline{l}inear \underline{c}omplementary \underline{d}ual} (LCD) code. LCD codes have first been considered by Massey in
\cite{M}. The nowadays interest of LCP codes aroused from the fact that they can be used
 in protection against side channel and fault injection attacks \cite{B},
\cite{Bringer}, \cite{CG}. In this context the security of a linear complementary pair $(C,D)$  can be measured
by the {\it security parameter} $\min\{\dist(C), \dist(D^\perp)\}$.
Clearly, if $D = C^\perp$, then the security parameter for $(C,D)$ is $\dist(C)$.\\

Just recently, it has been shown in \cite{C18} that for linear
complementary pairs $(C,D)$ the codes $C$ and $D^\perp$ are
equivalent if $C$ and $D$ are both cyclic or $2$D cyclic codes under the
assumption that the characteristic of $K$ does not divide the
length. In \cite{G18}, this result has been extended to the case
that both $C$ and $D$ are $n$D cyclic for $n \in \N$. In both papers
the proof is rather complicated and formulated in the world of
polynomials.

Recall that an $n$D cyclic code is an ideal in the algebra
$$R_n = K[x_1, \ldots,x_n]/\langle x^{m_1} -1, \ldots, x^{m_n}-1 \rangle,$$
and that $R_n$ is isomorphic to the group algebra $KG$ where
$  G = C_{m_1} \times \cdots \times C_{m_n} $ with cyclic groups
 $C_{m_i}$  of order $m_i$. Thus the above mentioned results are results on ideals in abelian group algebras.\\

A linear code $C$ is called a {\it group code}, or $G$-code, if $C$
is a right ideal in a group algebra $$KG = \{ a=\sum_{g\in G}a_gg
\mid a_g \in K\}$$ where  $G$ is a finite group. The vector space
$KG\cong K^{|G|}$ with basis $g \in G$ serves as the ambient space
and the weight function is defined by $\wt(a) = |\{g \in G \mid a_g
\not= 0 \}|$ (which corresponds to the classical weight function via
the isomorphism $KG\cong K^{|G|}$). Note that $KG$ carries a
$K$-algebra structure via the multiplication in $G$. More precisely,
if
 $a=\sum_{g\in G}a_gg$ and $b=\sum_{g  \in G}b_g g$ are given, then
$$ ab = \sum_{g \in G} (\sum_{h \in G} a_hb_{h^{-1}g}) g.$$
 In this sense $n$D cyclic codes
 are  group codes for abelian groups $G$ and vice versa since a finite abelian group is the direct product of cyclic groups.\\

There is a natural $K$-linear anti-algebra automorphism $\hat{}:KG
\longrightarrow KG$ which is given by  $g \mapsto g^{-1}$ for $g \in
G$ (in the isomorphism $KG\cong K^{|G|}$, the automorphism \ $\hat{}$ \
corresponds to a permutation of the coordinates). Thus we may
associate to each $ a = \sum_{g \in G} a_gg \in KG$ the {\it
adjoint}
$\hat{a} = \sum_{g \in G} a_gg^{-1}$ and call $a$ self-adjoint if $a = \hat{a}$. \\

In addition, the group algebra $KG$ carries a symmetric non-degenerate
$G$-invariant bilinear form $\langle.\,,.\rangle$ which is defined by
$$
      \langle g,h \rangle = \left\{ \begin{array}{cl}
               1 & \mbox{if } g=h \\
               0 &  \mbox{otherwise.}
               \end{array}
               \right.
$$
Here $G$-invariance means that $ \langle ag,bg\rangle = \langle a,b
\rangle $ for all $ a,b \in KG$ and all $g \in G$. Via the
isomorphism $KG\cong K^{|G|}$, the above form corresponds to the
usual Euclidean inner product. With respect to this form we may
define the dual code $C^\perp$ of a group code $C \leq KG$ as usual
and say that $C$ is self-dual if $C = C^\perp$. Note that for a
group code $C$ the dual $C^\perp$ is a right ideal since for all $c
\in C, c^\perp \in C^\perp$ and $g \in G$ we have
$$ \langle c, c^\perp g \rangle = \langle c g^{-1}, c^\perp \rangle = 0.$$
Thus with $C$ the dual $C^\perp$ is a group code as well.\\

 In \cite{W} we classified completely group algebras
which contain  self-dual ideals. More precisely, a self-dual $G$-code
exists over the field $K$
if and only if $|G|$ and the characteristic of $K$ are even.
 In \cite{CW} we investigated LCD group codes and characterized them  via self-adjoint
idempotents $e^2 = e = \hat{e}$ in the group algebra $KG$.\\

In this short note we prove the following theorem which includes the above mentioned results of \cite{C18} and \cite{G18}.
Observe that we require no assumption on the characteristic of the field $K$.\\

\noindent {\bf Theorem.} Let $G$ be a finite  group. If $C \oplus D
= KG$ where $C$ and $D$ are $2$-sided ideals in $KG$, then $D$ is
uniquely determined by $C$ and $D^\perp$ is permutation equivalent to $C$. In particular $\dist(D^\perp) = \dist(C)$. \\

In order to prove the Theorem  we state some elementary facts from representation theory. \\

\noindent
{\bf Definition.}  {\rm If $M$ is  a right $KG$-module, then the dual vector space $M^* = \ho_K(M,K)$ becomes a right $KG$-module via
$$     m(\alpha g) = (mg^{-1})\alpha $$
where $m \in M,\alpha \in M^*$ and $g \in G$. With this action $M^*$ is called the {\it dual module} of $M$. Clearly,  $\dim\, M^* = \dim\, M.$} \\

\noindent
{\bf Lemma A}. {\rm (Okuyama-Tsushima, \cite{OT})} \label{L1} If $ e=e^2 \in KG$, then  $\hat{e}KG \cong eKG^*$. In particular,
$\dim \, \hat{e}KG =\dim \, eKG$.\\[0.5ex]
{\bf Proof:} A short proof is  given in Lemma 2.3 of \cite{CW}. \hfill $\Box$\\

\noindent
{\bf Lemma B.}  If   $D=e KG$ with $e^2=e \in KG$, then $D^\perp =(1-\hat{e})KG$.\\[0.5ex]
{\bf Proof:} First observe that $\hat{e}^2 =\hat{e}$ and that $\langle ab,c \rangle = \langle b, \hat{a}c \rangle$ for all
$a,b,c \in KG$.
Thus
$$ \langle ea,(1-\hat{e})b \rangle = \langle e^2a, (1-\hat{e})b \rangle = \langle ea,\hat{e}(1-\hat{e})b \rangle =0$$
for all $a,b \in KG$. Hence, $(1-\hat{e})KG \subseteq D^\perp$. As
$$\begin{array}{rcl}
  \dim\, (1-\hat{e})KG & = & |G| - \dim \, \hat{e} KG \\
              & = & |G| - \dim\, eKG^* \qquad \text{(by Lemma A)} \\
                        & = & |G| - \dim \, eKG \\
                        & = & \dim\,  D^\perp
\end{array} $$
we obtain $(1-\hat{e})KG = D^\perp$. \hfill $\Box$ \\

\noindent
{\bf Proof of the Theorem:} Since $C \oplus D = KG$ we may write $D=eKG$ and $C=(1-e)KG$ for
a suitable  central idempotent
$$e=e^2 \in Z(KG)= \{a\mid ab=ba \ \text{for all} \ b \in KG \}$$ (see \cite{CW}).
Lemma B says that $D^\perp = (1-\hat{e})KG$. Via the map $\hat{}:KG \longrightarrow KG$ the $K$-linear code
$(1-\hat{e})KG$ is permutation equivalent to the $K$-linear code $KG(1-e)$. But $KG(1-e) = (1-e)KG=C$
 since $e$ is central, which  completes the proof. \hfill $\Box$ \\

If $C$ and $D$ are only right ideals, then $D$ is uniquely
determined by $C$, but $D^\perp$, in general, is not necessarily
permutation equivalent to $C$. It even may happen
that $\dist(D^\perp) \not= \dist(C)$ as the next example shows.\\

\noindent {\bf Example.} Let $K = \F_2$ and let
$$G =\langle a,b \mid a^7=1=b^2, a^b=a^{-1} \rangle $$
be a dihedral group of order 14. If we put
$$ e = 1+a + a^2 +a^4 +b +a^2b +a^5b +a^6b, $$
then $e =e^2$. With {\sc Magma} one  easily computes
$\dist((1-e)KG)=2$ and $\dist(KG(1-e))=3.$

Now let $C=(1-e)KG$ and $D=eKG$. By Lemma B, we have $D^\perp =
(1-\hat{e})KG$. Thus
$$ \dist(D^\perp)= \dist((1-\hat{e})KG) = \dist(KG(1-e)) = 3, $$
but
$ \dist(C) = \dist((1-e)KG) = 2. $ We like to mention here that $C$ and $D$ are quasi-cyclic codes. \hfill $\Box$ \\

\noindent
{\bf Remark.}
Let $|K|=q^2$. In this case we may consider the Hermitian inner product on $KG$ which is defined by
$$ \langle \sum_{g \in G} a_gg, \sum_{h \in G}b_hh \rangle = \sum_{g \in G} a_gb_g^q.$$
For $a =\sum_{g \in G} a_g g$ we put $a^{(q)} = \sum_{g \in G} a_g^qg$. With this notation we have
$$D^\perp = (1-\widehat{e^{(q)}})KG$$ in Lemma B. Applying the anti-automorphism $\hat{}:KG \longrightarrow KG$
we see that $D^\perp$ is permutation equivalent to $KG(1-{e^{(q)}})$. If in addition $e$ is central, then $e^{(q)}$ is central.
Thus
$D^\perp$ is permutation equivalent to $(1-e^{(q)})KG = ((1-e)KG)^{(q)}$.

It follow that
$$ \dist(D^\perp) = \dist((1-e)KG)^{(q)}) =\dist((1-e)KG) =\dist(C).$$
Thus, in the Hermitian case  a linear complementary pair $(C,D)$ of $2$-sided group codes $C$ and $D$  also has security parameter $\dist(C)$.


\begin{thebibliography}{lllll}

\bibitem{B} {\sc S.~Bhasin, J.-L.~Danger, S.~Guilley, Z.~Najim and X.T.~Ngo}, ``Linear complementary dual code improvement to strengthen
encoded circuit against hardware Trojan horses.'' In Proc.
\emph{IEEE Int. Symp. Hardware Oriented Secur. Trust} (HOST) 2015,
pp. 82-87.

\bibitem{Bringer} {\sc J.~Bringer, C.~Carlet, H.~Cabanne, S.~Guilley and H.~Maghrebi}, ``Orthogonal direct sum making: A smartcard friendly
computation paradigm in a code, with builtin protection against side-channel and fault attacks.'' in \emph{Proc. WIST}, Springer 2014,
pp. 40-56.

\bibitem{C18} {\sc C.~Carlet, C.~G\"uneri, F.~\"Ozbudak, B.~\"Ozkaya and P.~Sol\'e}, ``On linear complementary pairs of codes.''
\emph{IEEE Trans. Inform. Theory}, vol.~64, pp. 6583-6589, 2018.


\bibitem{CG} {\sc C.~Carlet} and {\sc S.~Guilley}, ``Complementary Dual Codes for Counter-measures
to Side-Channel Attacks. 
In ``Coding Theory and Applications.'' Eds. R.~Pinto, P. Rocha Malonek and P. Vettory, \emph{CIM Series in Math. Sciences}.
vol.~3, pp. 97-105, Springer 2015.

\bibitem{CW} {\sc J.~de la Cruz and W.~Willems}, `` On group codes with complementary duals.'' \emph{Des. Codes Cryptogr.},
vol.~86, pp. 2065-2073, 2018.


\bibitem{G18} {\sc C.~ G\"uneri, B.~\"Ozkaya and S.~Sayici}, ``On linear complementary pair of nD cyc\-lic codes.''
\emph{IEEE Commun. Lett.}, vol.~22, pp. 2404-2406, 2018.


\bibitem{M} {\sc J.L.~Massey}, ``Linear codes with complementary duals.''   \emph{Discrete Math.}, vol.~106/107, 337-342, 1992.

\bibitem{OT} {\sc T.~Okuyama and Y.~Tsushima}, ``On a conjecture of P. Landrock.'' \emph{J. of Algebra}, vol. 104, pp. 203-208, 1986.


\bibitem{W} {\sc W.~Willems}, ``A note on self-dual group codes.'' \emph{IEEE Trans. Inf. Theory}, vol.~ 48, pp. 3107-3109, 2002.



\end{thebibliography}
\end{document}